\documentclass[aps,prf]{revtex4-1}

\usepackage{color}
\usepackage{mathtools}
\usepackage{latexsym}
\usepackage{graphicx}
\usepackage{float}
\usepackage{dcolumn}
\usepackage{bm}
\usepackage{amssymb}
\usepackage{amsmath}
\usepackage{mathtools}
\usepackage[space]{grffile}
\usepackage{xcolor}

\usepackage[colorinlistoftodos]{todonotes}

\newcommand{\p}{\partial}

\newcommand{\be}{\begin{equation}}
\newcommand{\ee}{\end{equation}}
\newcommand{\bea}{\begin{eqnarray}}
\newcommand{\eea}{\end{eqnarray}}

\begin{document}


\title{Chiral propulsion: the method of effective boundary conditions}

\author{Leonid Korneev$^{1}$}
\author{Dmitri E. Kharzeev$^{2,3}$}
\author{Alexandre G. Abanov$^{1,4}$}
\affiliation{
    $^{1}$Department of Physics and Astronomy, Stony Brook University, Stony Brook, NY 11794, USA \\
    $^{2}$Center for Nuclear Theory, Department of Physics and Astronomy, Stony Brook University, Stony Brook, NY 11794, USA \\
    $^{3}$Department of Physics, Brookhaven National Laboratory, Upton, NY 11973-5000, USA \\
    $^{4}$Simons Center for Geometry and Physics, Stony Brook, NY 11794, USA
}

\date{\today}

\begin{abstract}
We propose to apply an ``effective boundary condition" method to the problem of chiral propulsion. For the case of a rotating helix moving through a fluid at a low Reynolds number, the method amounts to replacing the original helix (in the limit of small pitch) by a cylinder, but with a special kind of partial slip boundary conditions replacing the non-slip boundary conditions on the original helix. These boundary conditions are constructed to reproduce far-field velocities of the original problem, and are defined by a few parameters (slipping lengths) that can be extracted from a problem in planar rather than cylindrical geometry. We derive the chiral propulsion coefficients for spirals, helicoids, helically modulated cylinders, and some of their generalizations using the introduced method. In the case of spirals, we compare our results with the ones derived by Lighthill and find a very good agreement. The proposed method is general and can be applied to any helical shape in the limit of a small pitch. We have established that for a broad class of helical surfaces the dependence of the chiral propulsion on the helical angle $\theta$ is universal, $\chi\sim \cos\theta\sin 2\theta$ with the maximal propulsion achieved at the universal angle $\theta_m = \tan^{-1}(1/\sqrt{2})\approx 35.26^\circ$.

\end{abstract}
\maketitle

\tableofcontents

\section{Introduction}

The motion of a solid object in a viscous fluid is conventionally described by hydrodynamic Navier-Stokes equations with appropriate boundary conditions on the surface of the object. Typical boundary conditions used are the non-slip boundary conditions stating that the fluid is not moving relative to the object's surface. The problem is notoriously complicated because of the non-linearity of hydrodynamic equations resulting in turbulence. It is, however, dramatically simplified at low Reynolds numbers when nonlinear inertial terms in equations can be neglected. The regime of low Reynolds numbers is very relevant for micron-scale objects moving in the water and is of great interest, e.g., for modeling the motion of microorganisms \cite{Zhang,Ghosh,Schamel,Keaveny,walker,aif2018chiral,Purcell}. In this regime, the Navier-Stokes equation is reduced to a linear Stokes equation.

A problem of interest for this work is the transformation of the torque applied to solid objects into translational motion (propulsion) which occurs if the chiral symmetry - symmetry between left and right - is broken by the shape of an object or boundary conditions on its surface. We refer to this phenomenon as to a \textit{chiral propulsion}.\cite{aif2018chiral} A simple example of a chiral shape is given by a helicoidal shape characterized by symmetry w.r.t. the simultaneous rotation by angle $\phi$  and translation by $(\phi/2\pi)d$ along the same axis. We will refer to the length parameter $d$ of the helicoidal shape as the pitch. It is expected that the application of the torque along the symmetry axis of a helicoid will result in translational motion along that axis.  It turns out that even for this symmetry, it is hard to find a non-trivial shape for which Stokes equations can be solved exactly, and various approximations have been used to find a relation between the angular velocity and the translational propulsion velocity of a helicoidal system. 

Significant progress can be made for a spiral shape -- a helicoidal shape made out of a thin wire.  Lighthill, in his seminal work on a motion of a flagellum \cite{Lighthill} used the cross-section of the flagellum as a small parameter and related a swimming velocity of an arbitrary shape filament with a constant cross-section to forces acting on the filament. Singular solutions --  known as Stokeslet and source doublets -- placed along the centerline of the filament produce velocity field satisfying non-slip boundary conditions on the filament's surface in the leading order in the diameter of a cross-section. Using this approximate solution Lighthill was able to compute the velocity of the propulsion of a spiral made of a very thin wire \cite{Lighthill} (see Ref.~\onlinecite{Liu_flag} for review of this and other approaches).

The chiral propulsion has also been computed analytically for few other geometries (see Ref.~\onlinecite{li} for review). For example, the propulsion velocity of a rotating cylinder with a helical perturbation on its surface was calculated by using perturbation theory in the amplitude of deviations from a cylindrical base shape. Such perturbative approach successfully predicts the chiral propulsion of helicoidal structures with large pitch $d$ but fails in the case of a small pitch $d\ll R$, where $R$ is the radius of the cylinder.\cite{li} 

This work is based on a simple observation that for a small pitch $d\ll R$, the outer part of the helicoid looks like a surface of a cylinder modulated at a scale much smaller than the radius of the cylinder. This suggests replacing the actual fluid flow outside of that imaginary cylinder with a flow generated by effective boundary conditions on the corresponding imaginary surface. These boundary conditions are well-known Navier, or partial slip boundary conditions (PSBC). We propose to apply the method of effective PSBC to solving chiral propulsion problems in the limit of small pitch. We will check that the method produces the same results for the spiral as the ones obtained in Ref.~\onlinecite{Lighthill}. The proposed method can be used for other geometries as well.

The paper is organized as follows. In section II we state the chiral propulsion problem and analyze its geometric nature and symmetries. In section III we present the main results of the paper and compute the chiral propulsion using the proposed method of effective PSBC. We discuss possible applications in section IV and significance of our results in section V. Technical details are relegated to appendices.

\section{Chiral propulsion of an infinite helicoid}
    
\subsection{Chiral propulsion of a solid body moving through viscous fluid} 

Consider a solid body moving in an unbounded incompressible viscous fluid domain $\mathcal{M}$. We denote the surface of  the body by $\Gamma$ . The fluid incompressibility condition can be written as 
\begin{align}
    \bm\nabla\cdot\bm{v}=0\,, \qquad \bm{x}\in \mathcal{M}\,,
 \label{incompressibility}
\end{align}
where $\bm{v}(\bm{x})$ is the fluid velocity.
Since we are interested in the case of small Reynolds numbers we use the Stokes equation for the velocity of the fluid  
\begin{align}
	-{\bm \nabla}P+\eta\Delta {\bm v} = 0\,, \qquad \bm{x}\in \mathcal{M}\,.
 \label{N-S}
\end{align}
Here $\eta$ is the shear viscosity of the fluid, and $P$ is the pressure. The equation (\ref{N-S}) is assumed to be valid outside of the domain occupied by the solid body.

The non-slip boundary conditions on the surface require that the velocity of the fluid at the surface is equal to the velocity of the surface itself:
\begin{align}
	\bm{v}\Big|_\Gamma\ =\bm{U}+\bm\Omega\times\bm{X}\,,
 \label{eq:naive2}
\end{align}
Here $\bm{U},\;\bm\Omega$ are the velocity and the angular velocity of the rigid body, respectively. The vector $\bm{X}$ is the position vector of the element on the surface $\Gamma$.  We also require the vanishing of the fluid velocity at infinity
\begin{align}
    \bm{v}(\bm{x})\to 0\,, \quad \mbox{as}\quad  \bm{x}\to \infty\,.
 \label{bcinfinity}
\end{align}
The hydrodynamic equations (\ref{incompressibility}-\ref{bcinfinity}) fully determine the motion of the fluid given the shape of the solid body $\Gamma$ together with parameters $\bm{U},\bm{\Omega}$. However, often instead of the velocity $\bm{U}$ and angular velocity $\bm\Omega$ of the body we know the external force $\bm{F}_{ext}$ and torque $\bm\tau_{ext}$ acting on the body. 
The external force and torque acting on an object are 
\begin{align}
	F^{ext}_i &= -\int_{\Gamma} f_i({\bf X})\,d\sigma \,,
 \label{eq:Fext} \\
	\tau^{ext}_i &= -\epsilon^{ijk}\int_{\Gamma} X_j f_k({\bf X})\,d\sigma \,,
 \label{eq:tauext} 
\end{align}
where the integral is taken over the surface of the body, $\bm{X}$ is the position of the element of the surface. The stress force acting on the element $d\sigma$ of the surface is defined by the fluid motion and is given by
\begin{align}
    f_i = -P n_i +\eta (\partial_iv_j+\partial_j v_i)n_j ,
\end{align}
where $\bm{n}$ is a vector outward-normal to the surface. Here and in the following we will use the Einstein's convention of summation over repeated indices. 

As the Stokes problem is linear we expect the general linear relationship between the force and the torque acting on the body and linear velocity and angular velocity of the body:
\begin{align}
    \bm{F}^{ext}/\eta &=  -K \cdot\bm{U} -C^\dag \cdot\bm{\Omega}\,,
 \nonumber \\
    \bm{\tau}^{ext}/\eta &= -C \cdot\bm{U} - R \cdot\bm{\Omega} \,.
 \label{tensors}
\end{align}
Here $K,C,R$ are tensors characterizing the motion of a rigid body in an unbound fluid which depend only on the geometric shape of the body.\cite{Happel1983} \footnote{Notice that there is no dependence of $K, C, R$ on the viscosity coefficient, and these tensors are purely geometric. The Stokes phenomenon can invalidate this statement. In the presence of the Stokes phenomenon, the inertial terms of the Navier-Stokes equation cannot be uniformly neglected, and the dissipative viscosity could appear as a cutoff.} $K$ and $R$ are translation and rotation tensors, respectively. They are both symmetric. $C$ is the so-called coupling tensor which is not symmetric in general. $C$ represents a coupling between rotational and translational motion.

The problem of chiral propulsion can then be formulated in the following way. Consider a body moving through the fluid subject to the condition that the total external force acting on the body is zero\footnote{In a related approach, we could push the body through liquid with non-zero force and require zero-torque condition.}
\begin{align}
	\bm{F}^{ext} =0 \,.
 \label{eq:Fext0}   
\end{align}
Then from (\ref{tensors}) we immediately find that 
the velocity $\bm{U}$ and the angular velocity $\bm\Omega$ of the body are related by a linear relation
\begin{align}
	\bm{U} = \mathcal{X} \bm{\Omega}\,,\qquad \mathcal{X} =K^{-1}C^{\dag}\,.
 \label{Chi-tensor}
\end{align}
The tensor $\mathcal{X}_{ij}$ is fully determined by the shape of the solid body and does not depend on the history of the motion and other parameters such as shear viscosity $\eta$ as long as we are in the Stokes' regime. We refer to the tensor $\mathcal{X}_{ij}$ as the \emph{chiral propulsion tensor}. The chiral propulsion tensor is generally not symmetric $\mathcal{X}_{ij}\neq\mathcal{X}_{ji}$. 

\subsection{A helicoidal system} 

In this work, we are interested in the chiral propulsion of an infinite length body rotating around its helicoidal symmetry axis $z$. Let us first define what we mean by the helicoidal shape in general. 

Consider a cylindrical coordinate system $(r,\phi,z)$. We refer to the body as to an infinite length helicoidal system if it possesses mixed translational-rotational symmetry, i.e., transforms into itself under: \begin{align}
    (r,\phi,z)\to \left(r,\phi+\alpha, z+\frac{\alpha}{2\pi}d\right)\,.
 \label{mixed-symmetry}
\end{align}
Here the angle $\alpha$ is arbitrary, and $d$ is the pitch of the helicoidal system. \footnote{In mechanics this symmetry results in the conservation of the combination $\frac{d}{2\pi}P+M = const$, where $P$ is the total momentum of a system and $M$ is the total angular momentum \cite{LANDAU197613}.}

The symmetry (\ref{mixed-symmetry}) is also assumed to be a symmetry of the hydrodynamic boundary conditions on the surface of the helicoidal system. It is lower than the symmetry of Stokes equations (\ref{incompressibility},\ref{N-S}). In addition to breaking full translational and full rotational symmetries, a general helicoidal shape also breaks parity symmetry (reflection with respect to a plane containing $z$-axis).  

\begin{figure}[H]
  \centering
    \includegraphics[width=0.45\textwidth]{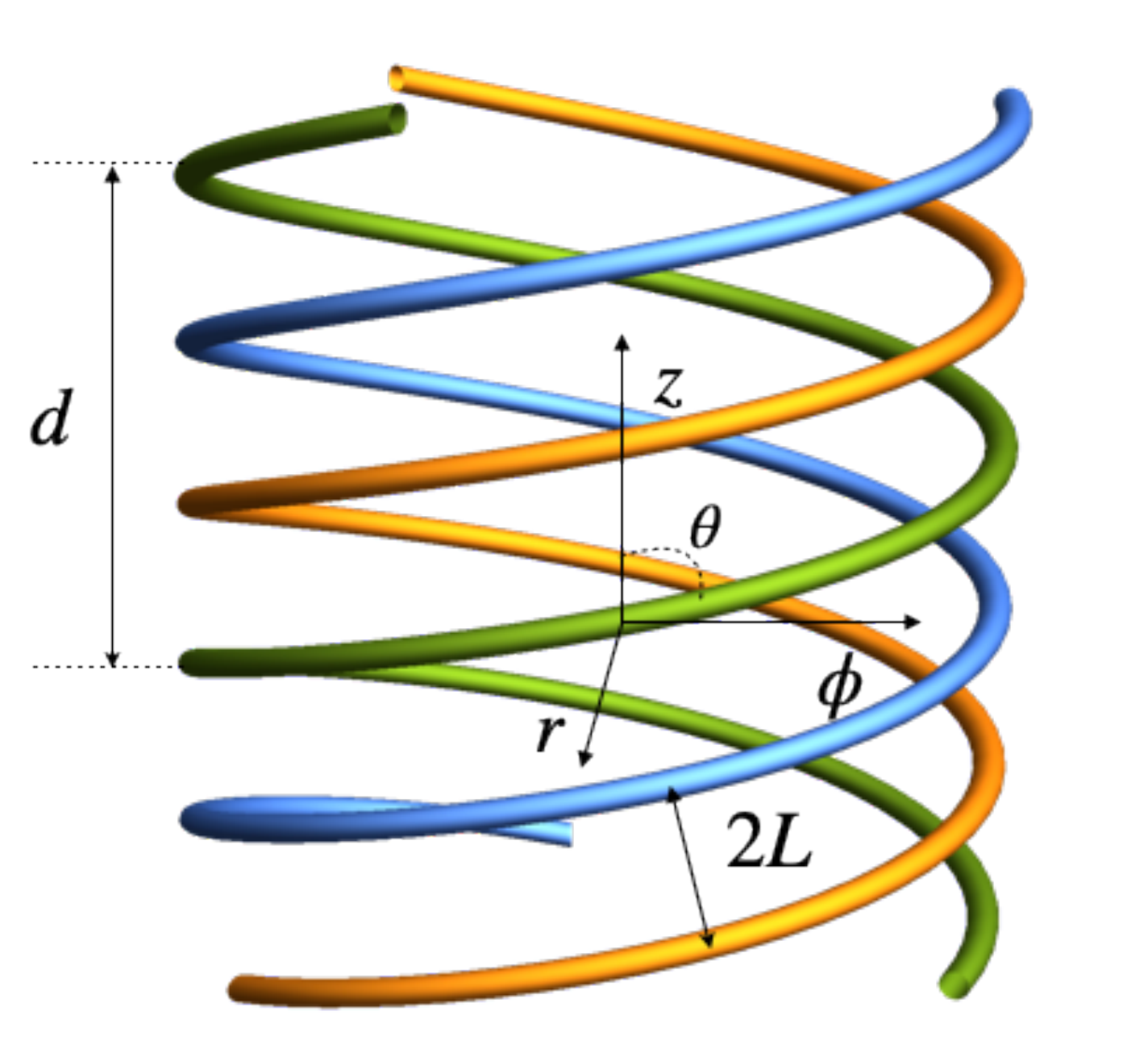}
  \caption{ Geometry of a $N$-spiral is shown for $N=3$. The angle $\theta$ between spirals and the $z$ axis is related to the pitch $d$ as $\cot\theta=\frac{d}{2\pi R}$. The distance $2L$ between nearest spirals is $2L=\frac{d}{N}\sin\theta$. }
  \label{Fig-N-spiral}
\end{figure}

In the following, we will also use shapes having an additional discrete symmetry 
\begin{align}
    (r,\phi,z)\to \left(r,\phi+\frac{2\pi}{N}, z\right)\,.
 \label{N-symmetry}
\end{align}
We refer to such shapes as $N$-helicoids. An example of such shape for $N=3$ is shown in Fig.\ref{Fig-N-spiral}. Assuming that the typical radius of the $N$-spiral is $R$ we have three length scales in the problem: $R,d,d/N$. It is also convenient to introduce the distance between spirals $2L$ and the angle $\theta$ formed by spirals with the axis of the helicoid:
\begin{align}
    \frac{L}{\pi R}=\frac{\cos\theta}{N}\,,
 \qquad 
    \tan \theta= \frac{2\pi R}{d} \equiv  \nu \,.
 \label{n-spiral-relations}
\end{align}
The inverse parameter $\nu^{-1}$ defined in (\ref{n-spiral-relations}) can be regarded as a dimensionless pitch of the helicoid. A limit of a small pitch corresponds to $\nu\to \infty$.

It is important to note that in the limit $N\to \infty$ the method of effective boundary conditions used in this work is exact.  Indeed, this method requires $2L\ll R$. Since $L \sim 1/N$, for sufficiently large $N$ the method applies for an arbitrary relation between $d$ and $R$.

\subsection{Chiral propulsion of a helicoid} 

Another consequence of this symmetry is that if the helicoid's angular velocity is aligned with the symmetry axis $z$, the propulsion velocity of the helicoid is also directed along the $z$-axis.  Then instead of the full chiral propulsion tensor $\mathcal{X}$ one has a \emph{chiral propulsion coefficient} $\chi$ defined as:
\begin{align}
	U_z =  \chi\,\Omega_z\,.
 \label{cp-coefficient}
\end{align}
The chiral propulsion coefficient is a pseudoscalar changing sign under parity (reflection with respect to a plane containing $z$-axis) as the velocity $U_z$ and angular velocity $\Omega_z$ transform as the components of a vector and pseudovector, respectively. As both $U_z$ and $\Omega_z$ are odd under time-reversal T, the chiral propulsion coefficient $\chi$ must be T-even and cannot depend linearly on the dissipative T-odd shear viscosity coefficient $\eta$. This symmetry argument agrees with the cancellation of the shear viscosity in (\ref{Chi-tensor}). Therefore, the effect of chiral propulsion is non-dissipative and geometric in nature.\cite{Purcell,shapere1987self,Note1} 

It is hard to find an exact solution for the Stokes problem for a helicoidal system. Therefore, in the following we use various approximations justified by  the smallness of additional parameters.

In the following we compute the chiral propulsion coefficients for shapes of helicoidal type using the method of effective boundary conditions. This method reduces the problem of computation of chiral propulsion for complex surfaces to the easier one for planar geometries.

\section{Small pitch limit. The method of effective boundary conditions}

Let us consider the limit of a small pitch of a helicoid. In this limit we will substitute the actual surface of the outer boundary of a helicoid by an effective cylindrical surface with modified boundary conditions. In this approach the derivation of chiral propulsion is elementary, with the difficulty shifted to the problem of finding a few parameters of the effective boundary conditions. 

\subsection{Effective boundary conditions for planar geometry}
\label{sec:ESPBCplanar}

We start by considering standard non-slip boundary conditions on the $z=0$ surface assuming that the fluid occupies the half-space $z>0$ . These boundary conditions state that the velocity of the flow is zero on $xy$-plane
\begin{align}
    \hat{\bm z}\times \bm{v} &=0\,, \quad \mbox{at } z=0\,.
 \label{noslip1}
\end{align}
Here $\hat{\bm z}$ is a unit vector normal to the surface, i.e., in the positive $z$-direction. In addition, we require that the fluid does not flow through the surface so that 
\begin{align}
    \hat{\bm z}\cdot \bm{v} &=0\,, \quad \mbox{at } z=0\,.
\end{align}
A solution of Stokes equations in the upper half-plane $z>0$ satisfying these boundary conditions is given by
\begin{align}
    \bm{v} &= \Omega z \hat{\bm e}\,,
\end{align}
where $\Omega$ is a constant defining the shear of the flow and $\hat{\bm e}$ is a unit vector parallel to the surface $\hat{\bm e}\cdot\hat{\bm n}=0$. 

Let us generalize the non-slip boundary condition (\ref{noslip1}) to 
\begin{align}
    v_i &= \tilde{h}_{ij}\sigma_{jz} \quad \mbox{at }z=0\,, \quad \mbox{for } i,j=x,y\,.
 \label{psbc1}
\end{align}
Here $\sigma_{jz}=\eta v_{jz}$ is a viscous shear stress tensor, $\eta$ is a shear viscosity and $v_{ij}$ is the strain rate defined by
\begin{align}
    v_{ij} &= \partial_i v_j+\partial_j v_i\,.
\end{align}
It is convenient to absorb the shear viscosity coefficient $\eta$ into the definition of the matrix $h_{ij}=\eta \tilde{h}_{ij}$ and write
\begin{align}
    v_i &=  h_{ij}v_{jz} \quad \mbox{at }z=0\,, \quad \mbox{for } i,j=x,y\,.
 \label{psbc2}
\end{align}

One can easily recognize in (\ref{psbc1}) the tensorial form \cite{bazant_vinogradova_2008} of Navier's partial slip boundary conditions (PSBC). \cite{navier1827lois} The matrix elements $h_{ij}$ have  dimensions of length defining the so-called slip lengths for a general case of an anisotropic surface. Physically, PSBC mean that there is a partial slip, i.e., finite velocity of the fluid along the surface proportional to the shear stress at the surface. The tensor of the second rank $h_{ij}$ is defined by the properties of the surface. 

Assuming that $h_{ij}$ is constant along the surface $z=0$ one can easily find the shear flow at $z>0$ satisfying (\ref{psbc1}). Namely, the solution for the flow is given by
\begin{align}
    v_z &=0\,,
 \\
    v_i &=  \Omega (z \hat{e}_i +h_{ij}\hat{e}_j)\,,
 \label{planarCP1}
\end{align}
where $\hat{e}_i$ are the components of a unit vector $\hat{\bm e}$ tangent to the surface. As it is seen from (\ref{planarCP1}), for a given direction $\hat{\bm e}$ of a shear flow at $z\to \infty$ there is a chiral (left or right) component of velocity perpendicular to the direction of the shear. This component $v_{ch}$ is given by
\begin{align}
    \chi &= \frac{v_{ch}}{\Omega} =  \hat{e}_i \epsilon_{ij}h_{jk}\hat{e}_k\,.
 \label{cpc10}
\end{align}
We will refer to the coefficient $\chi$ as to a chiral propulsion coefficient in planar geometry.
For a generic two by two real matrix $\hat{h}$ one can always find a reference frame such that the matrix has a form
\begin{align}
    \hat{h} &= \left(
    \begin{array}{cc}
        h_\perp & -h \\
        h & h_\parallel
    \end{array}\right)\,.
 \label{hmatrix1}
\end{align}
In this (principal) reference frame for a shear direction $\hat{\bm e}=(\cos\theta,\sin\theta)$ we find from (\ref{cpc10}) the chiral propulsion coefficient
\begin{align}
    \chi &= h + \frac{h_\parallel-h_\perp}{2}\sin 2\theta\,.
\end{align}
The matrix of effective slipping lengths (\ref{hmatrix1}) is also known as an interfacial mobility tensor. For passive surfaces, it has to be positive definite as a requirement that the surface does not transfer energy to the fluid, i.e., $h_\parallel, h_\perp\geq 0$ and $h^2\leq h_\perp h_\parallel$.  We refer the reader to Refs.~\onlinecite{bazant_vinogradova_2008,kamrin} for further details on the properties and applications of the tensorial slip boundary conditions.

We note that under the reflection with respect to, say, ``parallel'' axis $\theta\to -\theta$ we have $\chi\to-\chi$ and $h\to -h$, while $h_{\parallel,\perp}\to h_{\parallel,\perp}$. This means that the parameter $h$ of the Navier's tensor $h_{ij}$ is odd under parity and represents the parity violating parameters of the surface. 

For example, the plane modulated by identical grooves parallel to each other has a reflection symmetry with respect to at least one axis and, therefore, $h=0$ for that surface. In all examples considered in Section~\ref{sec:examples}, $h=0$; however, it is not prohibited to have non-zero $h$, and we keep it in our main equations for generality.

In the following, we will use the PSBC (\ref{psbc1}) as an effective boundary condition in the following way. Imagine a plane with grooves or some other modulations of its surface. Even if the actual boundary conditions at the surface are non-slip, at a distance $z$ from the surface such that $z$ is much larger than the characteristic length of surface modulations (e.g., the distance between grooves), the flow asymptotically takes the form of a shear\footnote{The next correction to the flow at large distances from the surface is expected to decay as $\exp(-kz)$, where $k=$ is a primary wavevector of the surface modulation.} (\ref{planarCP1}). This means that at large distances, the flow perturbed by the surface can be replaced by the one solving Stokes equation with effective PSBC (\ref{psbc1}). The actual shape of the surface will determine the parameters of the $h$-matrix $h_\parallel,\;h_\perp,\;h$. 

Before computing the PSBC parameters for few surfaces, let us present a computation of the chiral propulsion of the cylinder surface with PSBC (\ref{psbc1}).

\subsection{ Chiral propulsion of a cylinder with Navier boundary conditions}
 \label{cp-cyl}

Let us consider an infinite cylinder of radius $R$ having PSBC (\ref{psbc1}) on its surface. We assume that the cylinder rotates around its axis coinciding with the coordinate axis $z$ with angular velocity $\Omega$ and propels with velocity $U$ along the $z$-axis. We also assume that the effective matrix $h_{ij}$ is constant at the surface of the cylinder and has the form (\ref{hmatrix1}) in the local $xy$ coordinate system which is rotated by the angle $\theta$ clockwise with respect to the frame given by unit vectors $\hat{\bm \phi},\hat{\bm z}$ of cylindrical coordinate system $r,\phi,z$. Rotating (\ref{hmatrix1}) by an angle $\theta$ we obtain
\begin{align}
    \hat{h} &=  \left(
    \begin{array}{cc}
        h_{\phi\phi} & h_{\phi z} \\
        h_{z\phi} & h_{zz}
    \end{array}\right)
    =\left(
    \begin{array}{cc}
        h_\perp \cos^2\theta+h_\parallel \sin^2\theta & -h+\frac{h_\parallel-h_\perp}{2}\sin(2\theta) \\
        h+\frac{h_\parallel-h_\perp}{2}\sin(2\theta) & h_\perp \sin^2\theta +h_\parallel\cos^2\theta
    \end{array}\right)\,.
 \label{hmatrix10} 
\end{align}

On the surface of the cylinder we have from (\ref{psbc1})
\begin{align}
    ({\bm v}-\Omega R\hat{\bm\phi}-U\hat{\bm z})_i = h_{ij}\eta^{-1}\sigma_{jr}\,.
 \label{psbc10}
\end{align}
Here, in the l.h.s., we have a fluid velocity relative to the surface of the cylinder; the indices take values $i,j=\phi,z$ and the normal direction to the surface of the cylinder is $r$. The flow outside of the cylinder $r>R$ is given by the solution of the Stokes equation as:
\begin{align}
    v_r =0\,, 
 \qquad
    v_\phi = \frac{\Gamma}{2\pi r}\,,
 \qquad
    v_z = 0\,.
\end{align}
Using the expression for the stress tensor in cylindrical coordinates \cite{LL6}
\begin{align}
    \sigma_{\phi r} &= \eta\left(\frac{1}{r}\frac{\partial v_r}{\partial\phi}+\frac{\partial v_\phi}{\partial r}-\frac{v_\phi}{r}\right)\,,
 \\
    \sigma_{z r} &= \eta\left(\frac{\partial v_z}{\partial r}+\frac{\partial v_r}{\partial z}\right)\,,
 \\
    \sigma_{z z} &= 2\eta \frac{\partial v_z}{\partial z}\,,
\end{align}
we obtain from (\ref{psbc10})
\begin{align}
    \frac{\Gamma}{2\pi R}-\Omega R &= - h_{\phi \phi}\frac{\Gamma}{\pi R^2} \,,
 \\
    -U &= -h_{z \phi}\frac{\Gamma}{\pi R^2} \,.
\end{align}
This system gives both the chiral propulsion velocity $U$ and the circulation of the flow at infinity $\Gamma$ in terms of the angular velocity $\Omega$:
\begin{align}
    \frac{U}{\Omega R} &= \frac{2 h_{z\phi}}{R+2 h_{\phi\phi}}\,,
 \label{U10} \\
    \frac{\Gamma}{\Omega R^2} &= \frac{2\pi R}{R+2 h_{\phi \phi}}\,.
 \label{Gamma10}
\end{align}
The relation (\ref{U10}) and values (\ref{hmatrix10}) give for the chiral propulsion coefficient defined by (\ref{cp-coefficient}) the expression:
\begin{align}
    \frac{\chi}{R} &= \frac{2h+  (h_{\parallel}-h_{\perp})\sin 2 \theta }{R+ (h_{\parallel}+h_{\perp})- (h_{\parallel}-h_{\perp})\cos 2 \theta }\,.
 \label{cpmain}
\end{align}
This expression is the main result of this work. It gives the chiral propulsion coefficient of a cylinder in terms of parameters of Navier PSBC. In the following we apply (\ref{cpmain}) to compute the chiral propulsion coefficient for a few particular helicoidal systems. 

The chiral propulsion coefficient given by (\ref{cpmain}) is exact for a cylinder with Navier boundary conditions. However, to use these formulas for the realistic surface with \textit{effective} Navier boundary conditions, the slip lengths $h_{i j}$ and the period of modulation $2L$ must be much less than the typical radius of the cylinder, i.e.,  
\begin{align}
    h_{\perp},\; h_{\parallel},\; h,\; 2L \ll R\,.
 \label{applicab_cond}
\end{align}

\section{Examples and applications}
 \label{sec:examples}

In this section, we present several applications of (\ref{cpmain}) to the computation of the chiral propulsion coefficient for a few helicoidal shapes. 

\subsection{Example 1: Chiral propulsion of $N$-spiral}
 \label{sec:ex1}
 
Let us start with the geometry of a thin spiral and its generalizations. The chiral propulsion coefficient of an infinite spiral made out of a thin wire was computed by Lighthill \cite{Lighthill} using slender-body theory (see also Ref.~\onlinecite{Liu_flag}). In this section, we use a similar technique combined with the method of effective boundary conditions to compute the chiral propulsion coefficient and compare it with the one of Ref.~\onlinecite{Lighthill}.

\begin{figure}[H]
  \centering
    \includegraphics[width=0.7\textwidth]{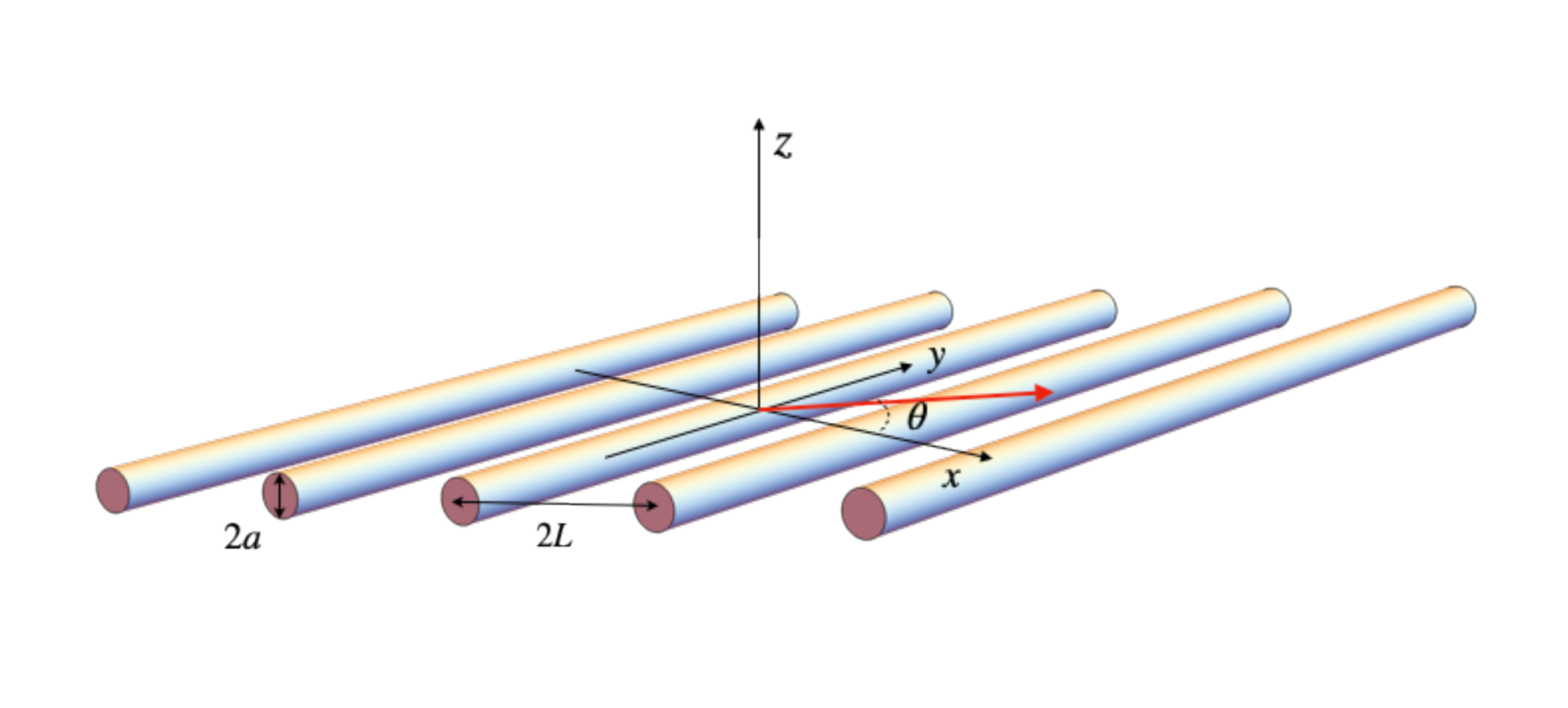}
  \caption{ An array of parallel wires with radius $a$ and the period $2L$. The red arrow points in the $\theta$ direction. }
  \label{fig:parwires}
\end{figure}

We consider the array of thin parallel wires placed in the shear flow shown in Fig.~\ref{fig:parwires}. The diameter of the wire $2a$ is assumed to be much smaller than the inter-wire distance $2a\ll 2L$.  We derive the following values for Navier parameters (see Appendix \ref{app:parwires} for details): 
\begin{align}
    \frac{h_{\perp}}{L} &= \frac{1}{2\pi}\left(-\frac{1}{2}+\ln{\frac{L}{\pi a}}\right)+O\left(\frac{a^2}{L^2}\right)\,, 
\label{hwires} \\
    \frac{h_{\parallel}}{L} &= \frac{1}{\pi}\ln{\frac{L}{\pi a}}+O\left(\frac{a^2}{L^2}\right)\,,
 \qquad h=0\,.
 \nonumber
\end{align}

Let us now consider a slight generalization of a spiral shape which we will refer to as $N$-spiral (see Fig.~\ref{Fig-N-spiral}). This shape is made out of $N$ infinite thin wires of radius $a$ with their centerlines satisfying:
\begin{align}
    r &= R\,, 
 \quad
    \frac{z}{d} = \frac{\phi}{2\pi} +\frac{k}{N}\,, 
 \quad 
    k=0,1,\ldots,N-1\,.
 \label{N-spiral}
\end{align}
Locally, the surface of the $N$-spiral looks like an array of parallel wires.  Assuming that the distance between wires is small, one can replace the $N$-spiral by a cylinder with Navier boundary conditions on its surface with parameters (\ref{hwires}) (see Appendix \ref{app:parwires} for the derivation). For the chiral propulsion of such a cylinder we obtain from (\ref{cpmain},\ref{hwires}) using (\ref{n-spiral-relations}): 
\begin{align}
    N\frac{\chi_N}{R} &= \frac{\left(\frac{1}{2}+\ln{\frac{L}{\pi a}}\right)\cos\theta \sin 2\theta}{2-\frac{L}{2\pi R}(1+\cos 2\theta) +\frac{L}{\pi R}\ln{\frac{L}{\pi a}}(3-\cos 2\theta)}\,.
 \label{cp-spiral-10}
\end{align}
Here $\chi_N$ is the chiral propulsion coefficient for $N$-spiral. There were two essential approximations used in deriving (\ref{cp-spiral-10}). One is of the order of $a^2/L^2$ when deriving the parameters (\ref{hwires}) for Navier boundary conditions. Another is the replacement of the spiral by the cylinder with effective slip boundary conditions. One expects the error of the latter approximation to scale as $h^{\perp,\parallel}/R$ or $L/R$.  

For $N=1$ the result (\ref{cp-spiral-10}) should be valid for $L/(\pi a)\gg 1$ and sufficiently small pitch $\nu\gg \ln\frac{L}{\pi a}$. In this limit $\theta\approx \frac{\pi}{2}-\frac{1}{\nu}$ and we further approximate (\ref{cp-spiral-10}) as:
\begin{align}
    \frac{\chi_1}{R} \approx \frac{L^2}{\pi^2 R^2}\frac{ \ln{(\frac{L}{\pi a})}+0.5}{1 + \frac{2L}{\pi R}\ln{(\frac{L}{\pi a}})}\,.
 \label{chi-spiral-Navier}
\end{align}
The Lighthill's result (\ref{LH-propulsion}) computed in the limit of a small pitch using (\ref{A1}-\ref{A3}) gives  
\begin{eqnarray}
    \frac{\chi_1}{R} \approx \frac{L^2}{\pi^2 R^2}\frac{ \ln{(\frac{L}{\pi a})}+0.5}{1 + \frac{2L}{\pi R}\ln{(\frac{L}{\pi a}})}\left(1-\frac{L}{\pi R}\right) \,.
 \label{chi-spiral-Lighthill}
\end{eqnarray}
It coincides with (\ref{chi-spiral-Navier}) to the leading order in $L/R$ as expected. 

Another interesting limit allowing for some analytic control is the limit of large $N$ for the propulsion of $N$-spiral. From the condition (\ref{applicab_cond}) and (\ref{hwires}) it follows that for sufficiently large $N\gg \frac{1}{2}\ln\frac{L}{\pi a}$ the result (\ref{cp-spiral-10}) should be valid for all pitches $\nu$ (i.e. for all angles $\theta$). 

Generalizing Lighthill approach we derive the chiral propulsion for the $N$-spiral  (Appendix~\ref{app:Nspiral}) without the use of effective boundary conditions and obtain (\ref{N-spiral-Lighthill}). This result should be valid for $\pi a\ll L$. On the other hand, for sufficiently large $N$ the effective boundary condition result (\ref{cp-spiral-10}) becomes applicable. This allows for a direct comparison of (\ref{N-spiral-Lighthill}) and (\ref{cp-spiral-10}). 

\begin{figure}[H]
  \centering
     \includegraphics[width=0.7\textwidth]{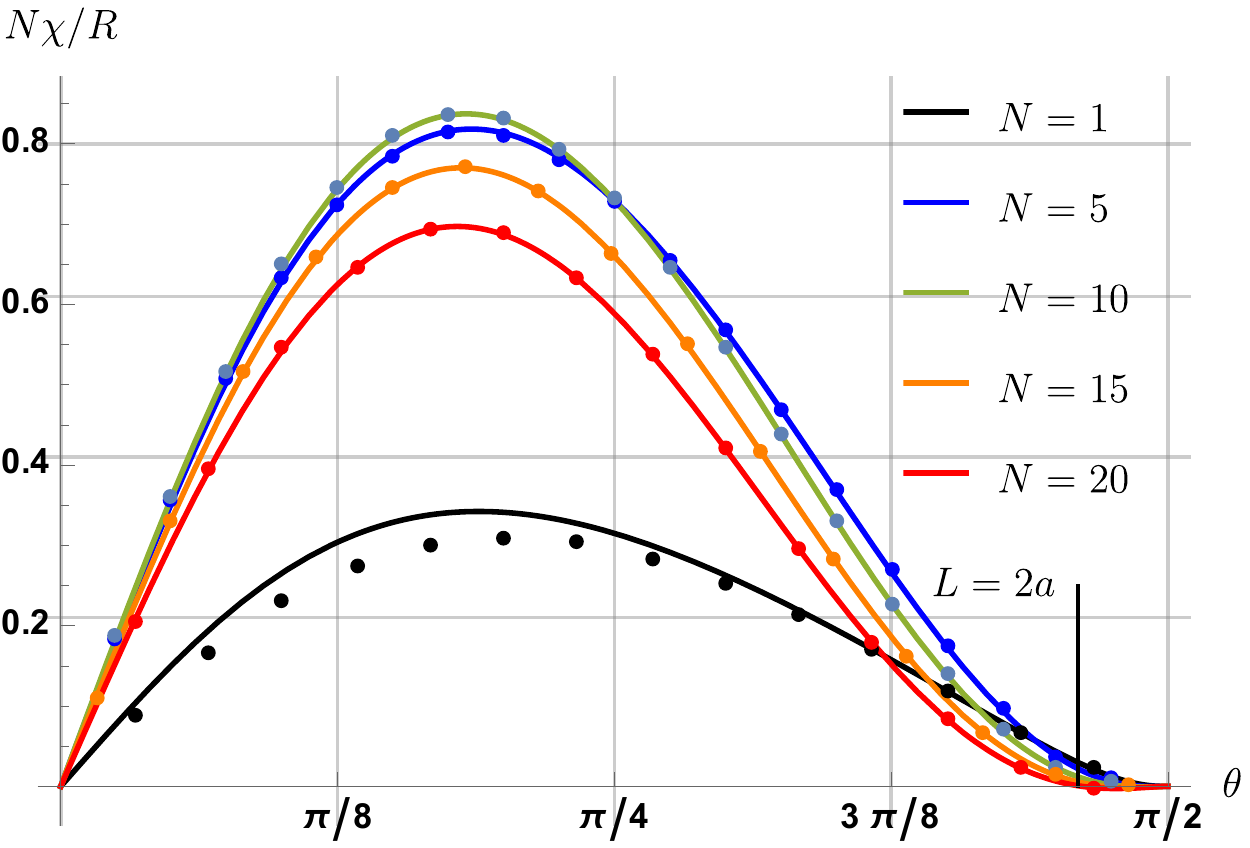}
  \caption{ The chiral propulsion coefficient for $N$-spiral with $N=1, 5, 10, 15, 20 $  and thickness $\frac{a}{R}=0.01$ is shown as $N\frac{\chi}{R}$ vs. $\theta$. The solid curves are given by (\ref{cp-spiral-10}), the dots are given by (\ref{N-spiral-Lighthill}) multiplied by N, the straight line indicates the pitch when the distance between wire centers $2L$ becomes twice their diameter $2a$ for N=20. }
  \label{Nspiral_comparison}
\end{figure}

We plot the chiral propulsion given by (\ref{cp-spiral-10}) (solid curves) and by the Lighthill-type formula (\ref{N-spiral-Lighthill}) (dots) as a function of the pitch parametrized by $\theta$ and at fixed $a/R=0.01$ in Fig.~\ref{Nspiral_comparison}. We see a very good agreement between the formulas already at $N=5$. When the pitch becomes very small, i.e., $\theta\to \pi/2$ the distance between wire centers $2L$ becomes comparable with wire diameters $2a$ and one should not rely on either of expressions (\ref{cp-spiral-10},\ref{N-spiral-Lighthill}).

\subsection{Example 2: Chiral propulsion of $N$-helicoid}
 \label{sec:ex2}
 
\begin{figure}[H]
  \centering
    \includegraphics[width=0.6\textwidth]{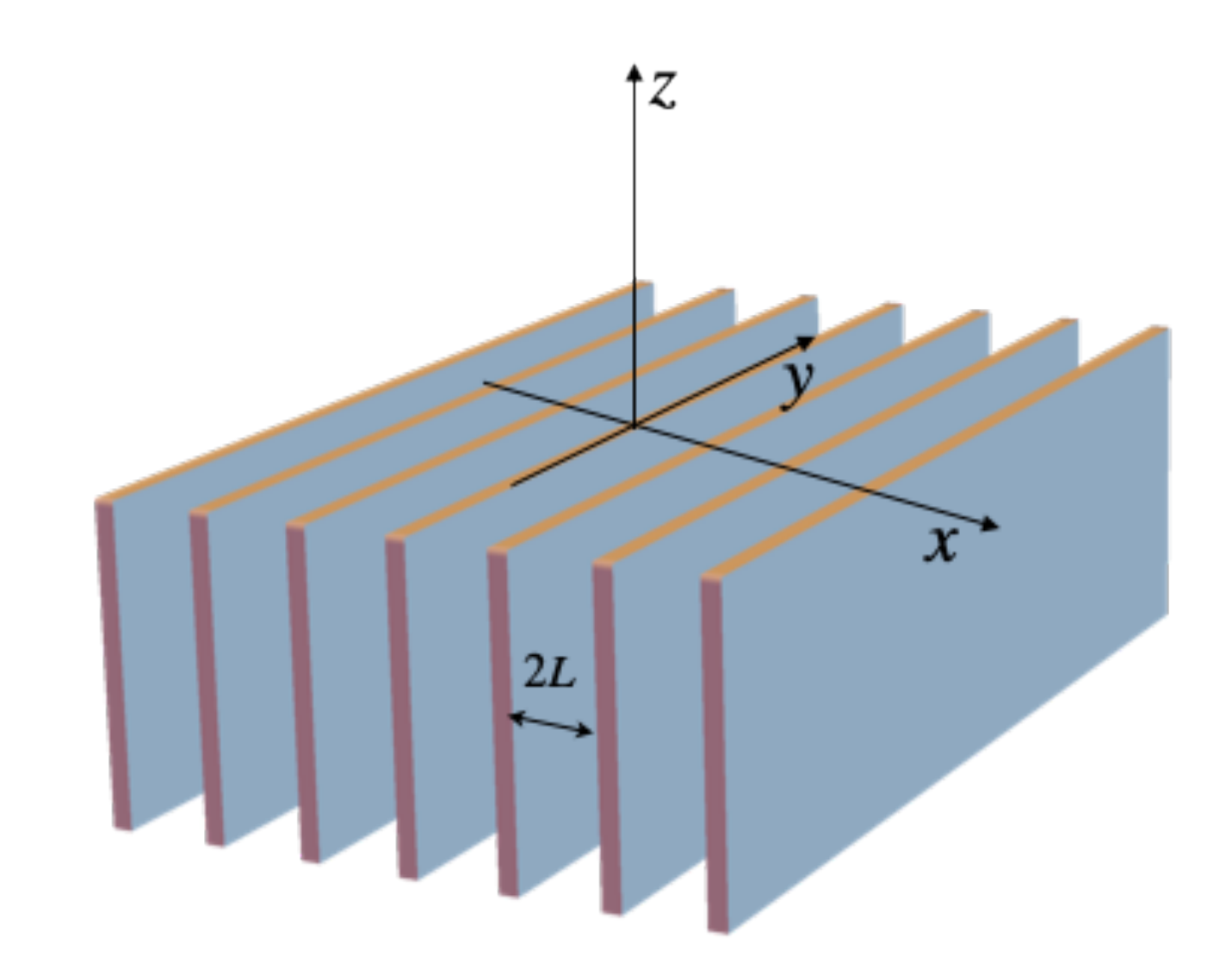}
  \caption{An array of protruding plates. Each plate is specified by $z\leq 0$ and $x=2Ln$, $n\in \mathbb{N}$. The thickness of plates is assumed to be vanishing.}
  \label{fig:plates}
\end{figure}

To the best of our knowledge, the closed-form solution of the chiral propulsion velocity of an infinite helicoid is not known. While it is technically possible to derive such a formula using the slender body theory, we do not attempt it in this work. Instead, we use the proposed method of effective boundary conditions to compute the chiral propulsion of a helicoid with a small pitch. We present it here as a particular case of a chiral propulsion velocity of $N$-helicoid, a generalization that is similar to the one we made for a spiral in the previous section.
We refer to the shape given in cylindrical coordinates by
\begin{align}
    r &\leq R\,, 
 \quad
    \frac{z}{d} = \frac{\phi}{2\pi} +\frac{k}{N}\,, 
 \quad 
    k=0,1,\ldots,N-1\,.
 \label{N-helicoid}
\end{align}
as $N$-helicoid. Here $\phi\in (-\infty,+\infty)$. The outer boundary of the $N$-helicoid (at $r=R$) is the previously introduced $N$-spiral (\ref{N-spiral}). The shape (\ref{N-helicoid}) has vanishing thickness and is essentially a surface. 
Locally a helicoid with a small pitch looks like an array of protruding plates (see Fig.~\ref{fig:plates}).
A planar geometry problem for this case is to find $h_{\perp}, h_{\parallel}$ for the system of semi-infinite plates given by equations
\begin{align}
    \Gamma=\Big\{z\leq 0,\;x=2Ln, \;\; n = 0,\pm1... \Big\}\,.
 \label{Gamma}
\end{align}
Exact values of $h_{\perp}, h_{\parallel}$ for this geometry can be found in literature \cite{davis,luchini} (see, e.g., equations 14 and 25 of Ref.~\cite{luchini}.). Here we give numerical values:
\begin{align}
    h_{\perp} &\approx 0.17 L\,,  \qquad
    h_{\parallel} 
    \approx 0.44 L\,,\qquad h=0\,.
 \label{hpar-barLsmall}
\end{align}

Replacing N-helicoid by a cylinder with Navier boundary conditions on its surface we compute the chiral propulsion coefficient using (\ref{cpmain}) and that $2L=\frac{2\pi R }{N}\cos{\theta}$:
\begin{align}
    N\frac{\chi_{N-helicoid}}{R}
    &\approx \frac{0.85 \cos\theta \sin{2\theta}}{1+N^{-1}\cos\theta(1.92-0.85\cos{2\theta})} \,.
 \label{cp-Nhelicoid-10}
\end{align}
The applicability of (\ref{cp-Nhelicoid-10}) is limited by the condition that both the distance between plates $2L$ and the corresponding slipping lengths $h_{\parallel,\perp}\sim L$ are much smaller than the radius of the helicoid. In this case the condition is simply $L\ll R$ or equivalently $\cos\theta/N\ll 1$. 
 
\begin{figure}[H]
  \centering
    \includegraphics[width=0.7\textwidth]{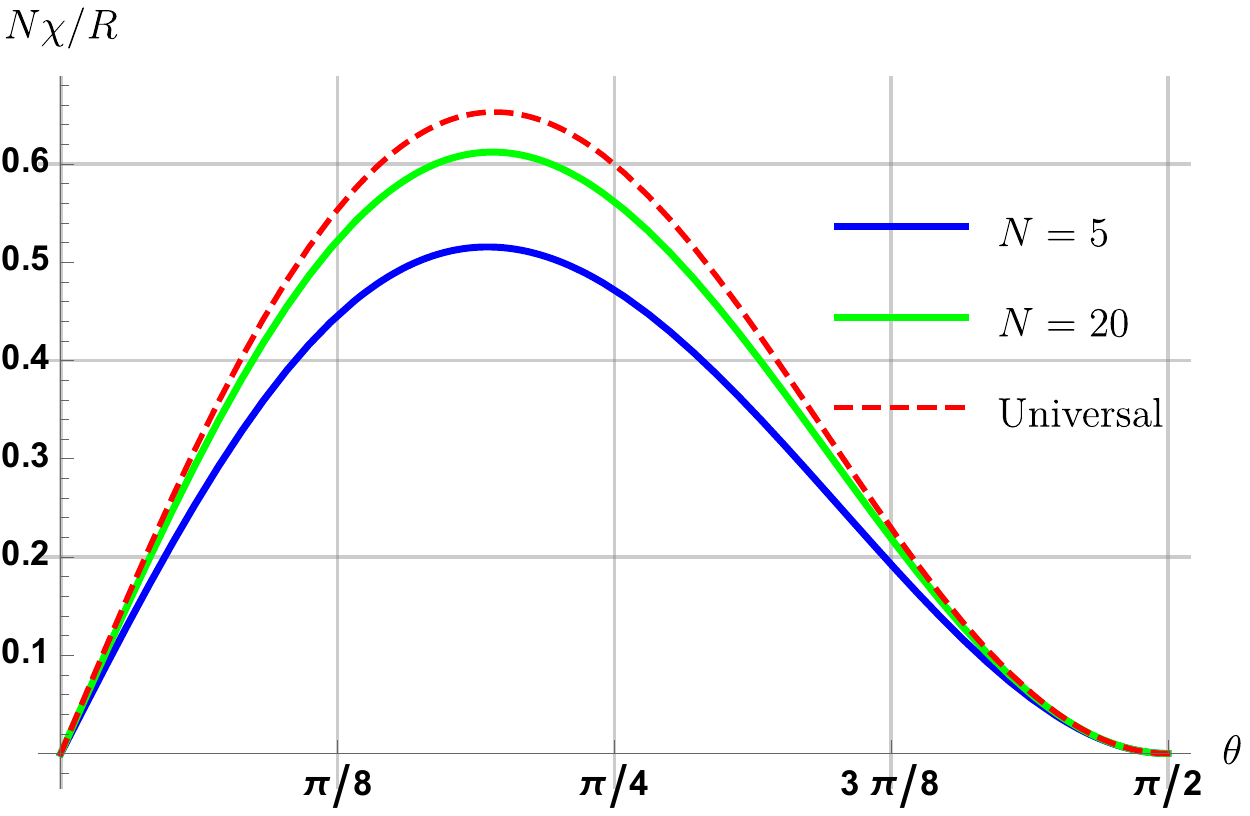}
  \caption{ The dependence of chiral propulsion coefficient for $N$-helicoid is shown as $N\frac{\chi}{R}$ vs. $\theta$ for N=5 and 20. The solid curves are given by (\ref{cp-Nhelicoid-10}). The dashed curve is given by (\ref{Nhelicoid}).}
  \label{fig:cpNhelicoid}
\end{figure}

In the limit of large $N$, neglecting subleading $1/N$  corrections, we obtain 
\begin{align}
    N\frac{\chi_{N-helicoid}}{R}\approx \, 0.85 \cos{\theta}\sin 2\theta\, ,
    \qquad  \cos{\theta}\ll N\,.
 \label{Nhelicoid}
\end{align}
The comparison of this formula with the more precise expression given by (\ref{cp-Nhelicoid-10}) is presented in Fig.~\ref{fig:cpNhelicoid}. For $N=20$ the slip lengths are about 10$\%$ of the radius and the chiral propulsion is within 10$\%$ of the universal formula (\ref{Nhelicoid}). 

Remarkably, the angle dependence of the chiral propulsion of the $N$-helicoid in the considered regime $L\ll R$ is universal and is given by Eq.~\ref{Nhelicoid}  (shown as a dashed red curve in Fig.~\ref{fig:cpNhelicoid}). It shows a maximum at $\theta_m=\tan^{-1}(1/\sqrt{2})\approx35.26^\circ\approx 0.196 \pi$.   This result is expected to be universal for a broad class of helical surfaces. The conditions for this are the applicability of the method of boundary conditions (surface details should be smaller than the object's radius) and that both $h_\perp$ and $h_\parallel$ are of the same order determined by the scale $L$. In this case, only the numeric coefficient in (\ref{Nhelicoid}) is surface-dependent in the leading order in $h_\perp/R$.

For the propulsion velocity of a helicoid (N=1) with a small pitch we obtain:
\begin{align}
    \frac{\chi_{helicoid}}{R}\approx 0.02 \frac{d^2}{R^2}\,,
    \qquad  d\ll R\,.
\label{helicoid}
\end{align}

Available experimental results for micro-swimmers with corkscrew shapes of various lengths\cite{walker,Schamel} exhibit velocity saturation when the length of swimmers becomes equal to few helical turns, so we can assume them to be of infinite length. However, pitches of the micro-swimmers in those experiments are typically larger than a helical radius, and our method does not apply.  A   naive application of Eq.~(\ref{cp-Nhelicoid-10}) to geometries considered in Refs.~\onlinecite{walker,Schamel} produces chiral propulsion  which is about twice the observed values. This disagreement is not surprising, as for large pitches considered in those references, the approximation of a constant force acting on their surface\cite{aif2018chiral} might be more applicable. The method of effective boundary conditions presented here is applicable only when force is concentrated at the edges of a helicoidal swimmer. 

\subsection{Example 3: Chiral propulsion of a modulated cylinder}
 \label{sec:ex3}
 
Now we consider the propulsion of a helical cylinder which cross section slightly deviates from a circle. The surface equation in cylindrical coordinates is given by
\begin{align}
   r(\phi,z)=R+Hf(\phi-2\pi z/d) \,,
\label{helicoidsurf}
\end{align}
where $f(\phi)$ is a $2\pi$-periodic function and the dimensionless deformation amplitude $\varepsilon \equiv H/R \ll 1$ quantifies the smallness of the deviation of the cross section from the circle.
 
When the pitch $d$ is much smaller than the radius or $f(\phi)$ has a period $2\pi/N$ with $N\gg 1$ the surface can be locally viewed as a periodically modulated plane with the height profile $y(x,z)=H f((x-\nu z)R^{-1})$ in planar geometry. Here the coordinates $x,y,z$ correspond to the directions $\phi,r,z$ in cylindrical geometry, respectively. For slightly deformed surfaces with an amplitude $\varepsilon$ and a period much larger than the amplitude $2\pi\nu^{-1}\gg \varepsilon$, we can use a domain perturbation theory and represent the velocity field as series in $\varepsilon$. Then the solution for velocity satisfying no-slip boundary conditions on the surface is obtained in each order in $\varepsilon$. In the second order in $\varepsilon$ (or $H$), the slipping lengths are given by \cite{luchini}:
\begin{align}
    h_{\perp} &=2h_{\parallel}=-4\frac{H^2}{R\cos\theta}\sum_{n=1}^\infty n |f_n|^2\,, \quad h=0\,,
 \label{hparmod}
\end{align}
where the Fourier coefficients of the modulation $f_n$ are defined as $f(\phi) = \sum_{n}  f_n e^{i n \phi }$, and we assumed $f_0=0$ without any loss of generality. When comparing (\ref{hparmod}) with the results of Refs.~\cite{luchini,kamrin} one should keep in mind that the wavevectors of the modulated plane are $k_n=n/(R\cos\theta)$ with $\tan\theta=\nu$. 

The relation $ h_{\perp}=2h_{\parallel}$ in (\ref{hparmod}) is specific for the second order in the amplitude of the modulation—the higher-order terms in amplitude break this relation. One might also expect the generation of the anti-symmetric part $h$ of the slip length matrix. However, it does not happen, at least in the third and fourth orders of the perturbation theory.  

Substituting (\ref{hparmod}) into (\ref{cpmain}) we obtain the chiral propulsion coefficient in the leading order in $h_{\parallel,\perp}/R$
\begin{align}
     \frac{\chi}{R} \approx  (h_\parallel-h_\perp)\sin{2\theta} =  4\varepsilon^2 \frac{\nu}{\sqrt{\nu^2+1}} \sum_{n=1}^\infty n|f_n|^2 \,.
     \label{planar_chiral_smooth}
\end{align}
This result reproduces  in the limit of small pitch $\nu \gg 1$ the result of  the computation by Li and Spagnolie \cite{li} made directly for the cylinder geometry:
\begin{align}
     \frac{\chi_{L-S}}{R} \approx  \left(1-\frac{3}{2\nu}\right)4\varepsilon^2 \sum_{n=1}^\infty n|f_n|^2\,.
 \label{LS_smooth}
\end{align}

Similarly to the case of the $N$-spiral considered in Section~\ref{sec:ex1}, the result (\ref{planar_chiral_smooth}) is based on two approximations. One is the basic approximation of the method of effective boundary conditions, namely $L\ll R$. The accuracy of the method is expected to be of the order of $L/R\sim 1/\nu$ and is supported by the comparison of (\ref{planar_chiral_smooth}) and (\ref{LS_smooth}). Another approximation involves the computation of slipping lengths for a given surface. Both (\ref{planar_chiral_smooth}) and (\ref{LS_smooth}) use the expansion in the amplitude of the modulation $H=\varepsilon R$ up to the second order. However, the actual dimensionless small parameter of the expansion is $H/L \sim N\varepsilon\nu$, where $2L$ is the wavelength of the modulation. \cite{li}

It is interesting to compare the numerical results obtained in Ref.~\onlinecite{li} to the prediction of the universal formula (\ref{cpmain}). The largest values $\nu=6$ and $N=4$ used in Ref.~\onlinecite{li} give $L/R\approx 1/8$ and make (\ref{cpmain}) applicable. However, the value $\varepsilon=0.1$ used in that reference for numerical computations translates into $H/L\sim N\varepsilon\nu \approx 2$, which is not small at all. Indeed, it was noticed in Ref.~\onlinecite{li} that the results of the perturbation theory extended to $\varepsilon^4$ while improving the agreement with numerics at small $\nu$ fail to describe the data at large $\nu$. 

In the following, we compare the numerics of Ref.~\onlinecite{li} with the predictions of the universal formula (\ref{cpmain}) used with the slipping lengths obtained numerically for planar geometry in Ref.~\onlinecite{luchini}.

Let us consider a simple cosine profile with amplitude $H$ and the wavelength $2L$, namely, $y(x,z) = H \cos(\pi x/L)$. On dimensional grounds we expect 
\begin{align}
    \frac{h_\parallel}{2L} &= - g_\parallel\left(s\right)\,,\qquad \frac{h_\perp}{2L} = - g_\perp\left(s\right) \,,
\end{align}
where $s=H/L$ is the dimensionless amplitude of the modulation. The asymptotics of the introduced functions $g_{\parallel,\perp}$ are given by
\begin{align}
    g_\parallel(s) &= \left\{
    \begin{array}{ll}
     \frac{\pi s^2}{4} & \mbox{for } s\ll 1 \,, \\
     \frac{s}{2}  -0.22  & \mbox{for } s\gg 1\,,
    \end{array}
    \right.
    \qquad
    g_\perp(s) = \left\{
    \begin{array}{ll}
     \frac{\pi s^2}{2} & \mbox{for } s\ll 1\,, \\
     \frac{s}{2} - 0.085 & \mbox{for } s\gg 1\,.
    \end{array}
    \right.
 \label{g-asymp}
\end{align}
Here the asymptotic at small $s$ is computed using the domain perturbation theory as in (\ref{hparmod}) (see Refs.~\onlinecite{luchini,kamrin}). The large $s$ limit corresponds to infinitely deep grooves $H\gg L$. In this limit, one can think of cosine modulation as a sequence of equally separated barriers and compute the asymptotics borrowing the result  (\ref{hpar-barLsmall}) corrected by the shift of the origin by $H$ to the maximums of barriers. \cite{luchini} In the limit of small pitch or large $N$ the small $s$ asymptotics (\ref{g-asymp}) used in the formula (\ref{cpmain}) should give a satisfactory result for chiral propulsion. However, it turns out that this approximation is not sufficient for the modulation amplitude values used in Ref.~\cite{li} and one needs more precise expressions for $g_{\parallel,\perp}$ interpolating between small and large $s$.

\begin{figure}[H]
  \centering
  \includegraphics[width=0.7\textwidth]{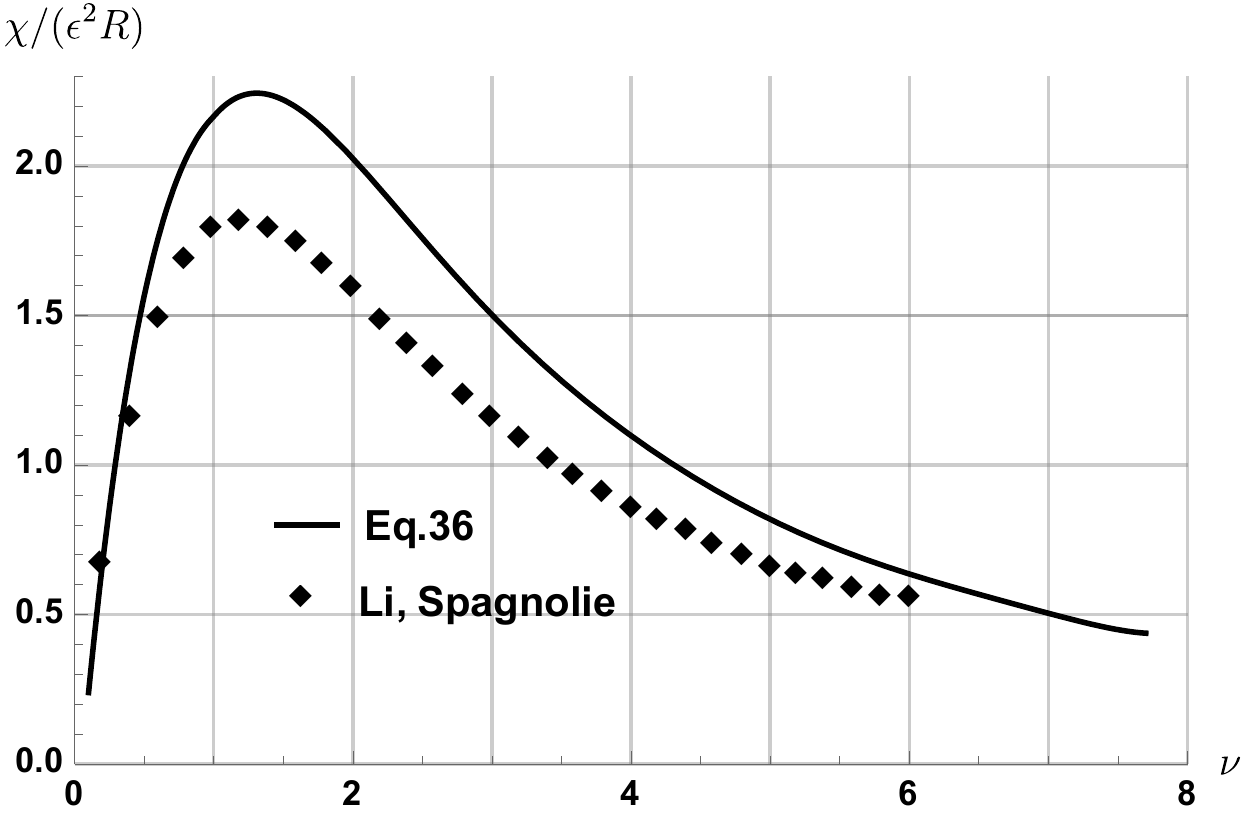}
  \caption{The dependence of chiral propulsion velocity is shown as $\frac{\chi}{\varepsilon^2 R}$ vs. $\nu$, where $\varepsilon=H/R=0.1$. The curve is given by (\ref{cpmain}) for the surface $f(\phi)=\cos({4\phi)}$. The slipping lengths $h_{\parallel,\perp}$ have been extracted from the Fig.~10 of Ref.~\cite{luchini}. The diamonds represent the $N=4$ data from the Figure 4b of Ref.~\onlinecite{li}.}
  \label{fig:cpNhelicoid1}
\end{figure}

Remarkably the functions $\bar{h}_{\parallel,\perp}\equiv s/2-g_{\parallel,\perp}(s)$ have been computed numerically and plotted for $s<1$ in Ref.~\onlinecite{luchini} (see Fig.~10 in that reference). We have extracted the functions $g_{\parallel,\perp}$ from the plot in Ref.~\onlinecite{luchini} and used them as $h_\parallel =-2L g_\parallel(\varepsilon R/L)$ with $2L=2\pi R\cos\theta/N$ and similarly for $h_\perp$ in our formula (\ref{cpmain}). The resulting plot of $\chi/R$ as a function of $\nu$ is shown for $\varepsilon=0.1$ and $N=4$ in Fig.~\ref{fig:cpNhelicoid1} on top of the numeric data extracted from Fig.~4b of Ref.~\onlinecite{li}.

Comparing Fig.~\ref{fig:cpNhelicoid1} with Fig.~4b of Ref.~\cite{li} we see that the solid curve produced by the method of effective boundary conditions (\ref{cpmain}) is in good agreement with numerical data for chiral propulsion in the whole range of $\nu$ for the modulation given by $f(\phi)=\cos N\phi$ with $N=4$. This agreement is to be compared with the Fig.~4b of Ref.~\cite{li} where it was shown that the small amplitude approximation fails in the small pitch (large $\nu$) regime.

We expect that (\ref{cpmain}) becomes asymptotically exact for $N\to\infty$. In this limit the curve will be taking a shape of a universal curve $\chi\sim \cos\theta\sin{2\theta} \sim \nu (\nu^2+1)^{-3/2}$ with the maximum at $\nu_m =\tan\theta_m=1/\sqrt{2}$ corresponding to $\theta_m\approx 35.26^\circ$. At $N=4$ the curve $\chi(\nu)$ shown in Fig.~\ref{fig:cpNhelicoid1} is still far from that universal limit and exhibits maximum at $\nu\approx 1.2$ or  $\theta\approx 50^\circ$.

\section{Conclusion}

To summarize, we applied the method of effective boundary conditions to the chiral propulsion of helical objects. The method is applicable in the limit of small pitch. In this limit, the chiral propulsion coefficient $\chi$ defined in Eq.~(\ref{cp-coefficient}) depends only on three slipping lengths entering as parameters $h_\parallel,h_\perp,h$ (\ref{hmatrix10}) of the effective Navier boundary conditions (\ref{psbc10}), or equivalently, the surface mobility tensor. The main result of this work is the formula (\ref{cpmain}) giving the chiral propulsion in terms of these parameters.

The result is universal and is valid for any object of helical symmetry in the limit of small pitch. More precisely, the method is self-consistent when the effective slipping lengths and surface modulation scale are much smaller than the typical radius of the helical object. The surface-specific $h$-parameters can be extracted from the solution of a simpler problem of shear flow over the surface, i.e., in the half-space geometry. We applied this method to compute chiral propulsion coefficients for $N$-spiral, $N$-helicoid, and a cylinder with a modulated surface in Section~\ref{sec:examples} with more technical parts relegated to appendices.   

We have established that for a broad class of helical surfaces (see the discussion after Eq.~\ref{Nhelicoid}), the dependence of the chiral propulsion on the helical angle $\theta$ is universal, $\chi\sim \cos\theta\sin 2\theta$ with the maximal propulsion achieved at the universal angle $\theta_m = \tan^{-1}(1/\sqrt{2})\approx 35.26^\circ$.

For the $N$-spiral shape, the chiral propulsion coefficient can be computed using the Lighthill approach. The original formula for a spiral with $N=1$ was derived in Ref.~\onlinecite{Lighthill} (see Appendix~\ref{app:spiral}). We presented its straightforward generalization for general $N$ in the Appendix~\ref{app:Nspiral}. We showed that in the limit of the small pitch, the universal formula (\ref{cpmain}) agrees with the more precise, surface-specific results (\ref{LH-propulsion},\ref{N-spiral-Lighthill}).

We used the existing numerical data and experiments performed for the propulsion of helical objects with a small pitch necessary to check the proposed chiral propulsion formula (\ref{cpmain}). The parameters entering (\ref{cpmain}) can be either independently computed or measured for a flow above the surface modulated in the same way as the surface of the helical object. In the small pitch regime, our universal formula (\ref{cpmain}) is in agreement with Lighthill's results (see Section~\ref{sec:ex1}) that have been thoroughly tested. \cite{Liu_flag} We also found a very good agreement of (\ref{cpmain}) with the numerical data of Ref.~\onlinecite{li} as discussed in Section~\ref{sec:ex3}.

{\sl Acknowledgements:}
The authors thank Gregory Falkovich for useful remarks. This work was supported by the US Department of Energy, Office of Science, Basic Energy Sciences, under Award No. 
US DOE DESC-0017662 (LK,DEK,AGA) 
and by the US National Science Foundation Grant NSF DMR-1606591 (AGA).


\appendix

\section{Chiral propulsion of spiral}
\label{app:spiral}

In his seminal paper Lighthill \cite{Lighthill} represented
a velocity field produced by a moving thin filament as an array of Stokeslets and source doublets. He applied this general method to derive the chiral propulsion of the spiral of radius $R$ and pitch $d$ defined by $\mathbf{r}(\phi)=  R(\cos{\phi}, \sin{\phi}, \epsilon \phi)$, $\phi\in(-\infty,+\infty)$ with $\epsilon=\frac{d}{2\pi R}=\cot\theta$ and obtained
\begin{align}
    \frac{\chi_1}{R} &= \frac{-\cos{\theta}\sin{\theta}+\cos{\theta}\csc^2{\theta} A_1 }{\cos^2{\theta}+\csc{\theta}(A_2+A_3)},
 \label{LH-propulsion}
\end{align}
where $A_{1,2,3}$ following Ref.~\onlinecite{Lighthill}  are given by 
\begin{align}
    A_1 &= \int_{\phi_c}^\infty d\phi\,\phi\sin\phi\,\xi^{-3/2}\,,
 \label{A1-def}\\
    A_2 &= \int_{\phi_c}^\infty d\phi\,\sin^2\phi \,\xi^{-3/2}\,,
 \label{A2-def}\\
    A_3 &= \int_{\phi_c}^\infty d\phi\,\cos\phi\,\xi^{-1/2}\,,
 \label{A3-def}
\end{align}
and
\begin{align}
    \xi &= \epsilon^2\phi^2+4\sin^2\frac{\phi}{2}\,.
 \label{xi-def}
\end{align}
These integrals are not independent but related by the following identity (can be easily derived by integrating (\ref{A3-def}) by parts)
\begin{eqnarray}
   \epsilon^2 A_1= A_3-A_2+\frac{\sin{\phi_c}}{\sqrt{4\sin^2{\frac{\phi_c}{2}}+\epsilon^2\phi_{c}^2}}\,.
 \label{A-relations}
\end{eqnarray}

The short distance cutoff used in the definition of integrals in (\ref{A1-def}-\ref{A3-def}) is given by 
\begin{align}
    \phi_c &=\frac{a\pi\sqrt{e}}{2\pi R\sqrt{1+\epsilon^2}}  \,.
 \label{phic1}
\end{align}
Here $e=2.718\ldots$ not to be confused with the parameter $\epsilon$. For a single spiral we use $2L=2\pi R\cos\theta$ as a shortest distance between neighboring turns and $\epsilon = \cot\theta$ to rewrite conveniently
\begin{align}
    \phi_c &=   \frac{\pi a}{L}\frac{\sqrt{e}}{2}\frac{\epsilon}{1+\epsilon^2}\,.
\end{align}

We compute the Lighthill integrals (\ref{A1-def}-\ref{A3-def}) in the limit of a small pitch $\epsilon\ll 1$  in Appendix~\ref{sec:Aintegrals}. Using the results (\ref{A1}-\ref{A3}) we obtain for the chiral propulsion (\ref{LH-propulsion})

\begin{align}
    \frac{\chi_1}{R}   
     &\approx\    \epsilon^2 \frac{1+ \ln(\frac{\epsilon}{2\phi_c})}{1+ 2\epsilon\ln(\frac{\epsilon}{2\phi_c})}\left(1-\epsilon
     +\frac{3\epsilon^2}{2}\frac{\ln{\phi_c}}{1+\ln{\frac{\epsilon}{2\phi_c}}}\right)
\end{align}

\section{Chiral propulsion of $N$-spiral}
\label{app:Nspiral}

It is straightforward to generalize the Lighthill's result (\ref{LH-propulsion}) for the propulsion of the $N$-spiral -- the shape of a thin filament defined by its centerline $\vec{r}(\phi)= R \left(\cos{\phi},\sin{\phi}, \epsilon\left(\phi-\phi_n\right)\right)$ where $ n=0,...,N-1,$, $\phi_n=2\pi\frac{n}{N}$, and $\phi\in(-\infty,+\infty)$. The $N$-spiral consists of $N$ identical copies of a single spiral arranged equidistantly along the common $z$-axis. Simply applying the Lighthill's derivation of (\ref{LH-propulsion}) for to the $N$-spiral we obtain
\begin{align}
    \frac{\chi_N}{R} = \frac{-\cos{\theta}\sin{\theta}+\cos{\theta}\csc^2{\theta} \sum_{n=0}^{N-1} B_{1,n} }{\cos^2{\theta}+\csc{\theta}\sum_{n=0}^{N-1} (B_{2,n}+B_{3,n})}\,,
 \label{N-spiral-Lighthill}
\end{align}
where $n=1,2,\ldots,N-1$, 
\begin{align}
    B_{1,n} &=\frac{1}{2} \int_{-\infty}^\infty d\phi\,\phi\sin\phi\,\xi_n^{-3/2}\,,
 \label{B1-def}\\
    B_{2,n} &= \frac{1}{2}\int_{-\infty}^\infty d\phi\,\sin^2\phi \,\xi_n^{-3/2}\,,
 \label{B2-def}\\
    B_{3,n} &= \frac{1}{2}\int_{-\infty}^\infty d\phi\,\cos\phi\,\xi_n^{-1/2}\,,
 \label{B3-def}
\end{align}
and
\begin{align}
    \xi_n &= \epsilon^2(\phi-\phi_n)^2 +4\sin^2\frac{\phi}{2}\,,
 \nonumber\\
    \phi_n &= 2\pi \frac{n}{N}\,.
\end{align}
We assume here that by definition $B_{1,0}=A_1$, $B_{2,0}=A_2$, and $B_{3,0}=A_3$ with the same cutoff scale $\phi_c$ as in (\ref{phic1}).

\section{A-integrals in the limit of small pitch}
 \label{sec:Aintegrals}

For $A_1$ integral we have
\begin{align}
    A_1 &= \int_{\phi_c}^{+\infty}d\phi\, \frac{\phi\sin\phi}{\left(4\sin^2\frac{\phi}{2}+\epsilon^2\phi^2\right)^{3/2}} = A_1^{\phi_c}+A_1'\,,
\end{align}
where for $\epsilon\ll 1$
\begin{align}
    A_1^{\phi_c} &= \int_{\phi_c}^{\pi}d\phi\, \frac{\phi\sin\phi}{\left(4\sin^2\frac{\phi}{2}+\epsilon^2\phi^2\right)^{3/2}}
 = \frac{1}{(1+\epsilon^2)^{3/2}}\ln\frac{\pi}{\phi_c} +1-\frac{\pi}{2} +\ln\frac{4}{\pi} +\ldots
 \nonumber \\
    &= 1-\frac{\pi}{2} +\ln\frac{4}{\phi_c} + \frac{3\epsilon^2}{2}\ln\phi_c +O\left(\phi_c^2,\epsilon^2\right)\,.
 \label{A1fc}
\end{align}
In the second line of (\ref{A1fc}) all terms containing $\ln(\phi_c)$ are included explicitly. 

The remaining part of the integral is given by
\begin{align}
    A_1' &= \sum_{n=1}^{\infty}\int_{-\pi}^{+\pi}d\phi\, \frac{(\phi+2\pi n)\sin\phi}{\left[4\sin^2\frac{\phi}{2}+\epsilon^2(\phi+2\pi n)^2\right]^{3/2}} \,,
\end{align}
where we replaced the integral over $\phi$ to the sum of integrals. 
Using Euler-Maclaurin formula we computed
\begin{align}
    A_1' &= \ln\frac{\epsilon}{8} +\frac{\pi}{2}+1 +C +O\left(\epsilon^2\ln\epsilon \right)\,.
\end{align}
Here the constant $C$ was obtained numerically and as it vanishes with a good accuracy we conjecture that $C=0$. In the following we use $C=0$.  

Combining the above we have for $A_1=A_1^{\phi_c}+A_1'$:
\begin{align}
    A_1\approx \ln{\frac{\epsilon}{2\phi_c}}+2+\frac{3\epsilon^2}{2}\ln{\phi_c} +O\left(\epsilon^2\ln \epsilon ,\phi^2_c\right) \,.
    \label{A1}
\end{align}

The calculation of $A_2$ and $A_3$ can be done in the same manner giving
\begin{align}
    A_2&\approx \frac{1}{2\epsilon}+\ln{\frac{\epsilon}{2\phi_c}}-1+\frac{3\epsilon^2}{2}\ln{\phi_c} +O\left(\epsilon^2\ln{\epsilon},\phi^2_c\right) \,,
   \label{A2} \\
     A_3&\approx \frac{1}{2\epsilon}+\ln{\frac{\epsilon}{2\phi_c}}-2+\frac{\epsilon^2}{2}\ln{\phi_c} +O\left(\epsilon^2\ln{\epsilon},\phi^2_c\right) \,.
    \label{A3}
\end{align}

\section{Streamlining the array of parallel wires}
\label{app:parwires}

Let us consider an infinite array of parallel round wires. We assume that wires are thin so that their radii $a$ are much smaller than the distance between wires $a\ll 2L$.  
We choose the coordinate system such that the centerlines of wires are given by equations:
\begin{align}
    z=0, \quad x=2nL,\; n=0,\pm 1, \pm 2, \ldots\,.
\end{align}
We assume no-slip boundary conditions on the surfaces of the wires. We will fix the shear flow at $z\to +\infty$ and assume that the flow velocity is constant as $z\to -\infty$. The corresponding solution of Stokes equation will determine the slipping lengths $h,h_\perp,h_\parallel$ as defined in Section~\ref{sec:ESPBCplanar}.

The symmetry of the problem immediately gives $h=0$. To determine $h_\parallel$ and $h_\perp$ we exploit the linearity of Stokes equations and solve two problems for the parallel and transverse shear flows separately. 

\subsection{Shear flow parallel to wires}

Let us assume that the shear flow at $z\to +\infty$ is parallel to the direction of the wires $y$-axis. We use the symmetry of the problem and look for the $y$-independent solution with $v_x=v_z=0$. In this case the Stokes equations (\ref{N-S}) can be reduced to the form
\begin{align}
    \Delta v_y = 0\,,
 \label{Lvy}
\end{align}
where $\Delta=\partial_x^2+\partial_z^2$ is the two-dimensional Laplace operator. 
This is the 2d Laplace equation that should be solved outside of wires with the following boundary conditions:
\begin{align}
    v_y &=0\,,\hspace{1.5cm} (x-2nL)^2+z^2=a^2,\;\; n=0,\pm1...
 \nonumber \\
    v_y &\to \Omega(z + h_{\parallel})\,,\hspace{2.2cm}   z \to +\infty
 \nonumber \\
    v_y &\to  \Omega h_1\,\hspace{3.3cm}  z\to -\infty
 \label{vyparallel-bc}
\end{align}

We will show that the following expression is the solution in the leading order in $a/L\ll 1$:
\begin{align}
    v_y &= \frac{\Omega}{2} z-\frac{\Omega}{2}\frac{\pi a^2}{2L}\frac{\sinh{\frac{\pi z}{L}}}{\cosh{\frac{\pi z}{L}}-\cos{\frac{\pi x}{L}}} 
    + \frac{\Omega L}{2\pi} \log\left(\cosh{\frac{\pi z}{L}}-\cos{\frac{\pi x}{L}}\right) 
    -\frac{\Omega L}{2\pi}\log{\frac{\pi^2 a^2}{2L^2}} \,,
 \nonumber \\
    \tilde{v}_y &= 2\tilde{z}+ \left[1-\tilde{a}^2\frac{\p}{\p \tilde{z}}\right] \log\left[ \left(\frac{\sinh\tilde{z}}{\tilde{a}}\right)^2+\left(\frac{\sin\tilde{x}}{\tilde{a}}\right)^2\right]\,,
 \nonumber \\
    \tilde{v}_y &= v_y\Big/\frac{\Omega L}{2\pi}\,, \quad \tilde{z} = \frac{\pi z}{2L}\,, \ldots \,. 
  \label{vyparallel-10}
\end{align}
Indeed, it is straightforward to check that it is the solution of the Laplace equation (\ref{Lvy}). In fact, every term of (\ref{vyparallel-10}) separately solves the Laplace equation. As the solution (\ref{vyparallel-10}) is $2L$-periodic with respect to $x$, it is sufficient to apply the no-slip boundary condition only on the surface of the wire at the origin, i.e., at $\Gamma:\;x^2+z^2=a^2$. Expanding in $z/a$ and $x/a$ we obtain from (\ref{vyparallel-10}):
\begin{align}
    \frac{v_y}{\Omega L}\Big|_{\Gamma} &= \frac{z}{2L} -\frac{1}{2L}\frac{\pi a^2}{2L}\frac{\pi z}{L}\frac{1}{\frac{\pi^2(z^2+x^2)}{2L^2}}\left(1+O\left(\frac{a^2}{L^2}\right)\right)
 +\frac{1}{2\pi}\left(\log\left(\frac{\pi^2(z^2+x^2)}{2L^2}\right)+O\left(\frac{a^2}{L^2}\right)-\log{\frac{\pi^2 a^2}{2L^2}}\right)
 = O\left(\frac{a^2}{L^2}\right)\,.
\end{align}
A typical velocity scale around wires is $\Omega L$ and we see that the solution (\ref{vyparallel-10}) satisfies no-slip boundary condition with $a^2/L^2$ accuracy.

In the limit $z\to -\infty$ we have up to exponentially decaying terms
\begin{align}
    \frac{v_y}{\Omega L}\Big|_{z\to-\infty} &\approx \frac{z}{2L}+\frac{\pi a^2}{4L^2}-\frac{1}{2\pi}\frac{\pi z}{L}+\frac{1}{2\pi}\log{2}-\frac{1}{2\pi}\log{\frac{\pi^2 a^2}{2L^2}}
\approx \frac{\pi a^2}{4L^2}-\frac{1}{\pi}\log{\frac{\pi a}{L}}
\end{align}
so that the shear is absent at $z\to -\infty$ and the last condition of (\ref{vyparallel-bc}) is satisfied with  
\begin{align}
    h_1=-\frac{L}{2\pi} \left(\log{\frac{\pi a}{L}}-\frac{\pi^2 a^2}{2 L^2}+O\left(\frac{a^4}{L^4}\right)\right)\,.
\end{align}

Similarly for $z\to+\infty$
\begin{align}
    \frac{v_y}{\Omega L}\Big|_{z\to+\infty} &\approx \frac{z}{2L}-\frac{\pi a^2}{4L^2}+\frac{1}{2\pi}\frac{\pi z}{L}+\frac{1}{2\pi}\log{2}+\log{\frac{\pi^2 a^2}{2L^2}}
 \approx  \frac{z}{L}-\frac{\pi a^2}{4L^2}-\frac{1}{\pi}\log{\frac{\pi a}{L}} \,.
\end{align}
Comparing to (\ref{vyparallel-bc}) we find the slipping length:
\begin{align}
    h_\parallel=\frac{L}{\pi} \left(\log{\frac{L}{\pi a}}-\frac{\pi^2  a^2}{4L^2}+O\left(\frac{a^4}{L^4}\right)\right)\,.
\end{align}

\subsection{Shear flow transverse to wires}

Now we consider the flow with the shear transverse to wires. Namely, we will look for a solution of Stokes equations satisfying $v_y=0$ everywhere and the following boundary conditions:
\begin{align}
    v_x &= v_z=0\,,\hspace{1.2cm} (x-2nL)^2+z^2=a^2,\;\; n=0,\pm1...
 \nonumber \\
    v_x &\to \Omega(z + h_{\perp})\,,\hspace{2.7cm}   z \to +\infty
 \nonumber \\
    v_x &\to  \Omega h_2\,\hspace{3.8cm}  z\to -\infty \,.
 \label{vperp-bc}
\end{align}

The problem is essentially two-dimensional. It is convenient to use a stream function defined as
\begin{align}
    v &= 2i\partial \psi\,,
\end{align}
where we used complex notations $v=v_x-iv_z$ and $\partial=\frac{\partial}{\partial w}$ with $w=x+iz$. In terms of the stream function the Stokes equation (\ref{N-S}) for two-dimensional flow reduces to the biharmonic equation $\Delta^2\psi=0$. Our goal is to find the function $\psi$ as an expansion in $a/L\ll 1$.
It can be checked that the following potential satisfies the boundary conditions (\ref{vperp-bc}) in leading orders in $a/L$
\begin{align}
    \psi &= \frac{\Omega}{4} z^2+Az +\psi_S +\psi_{SD}+\psi_D\,,
 \label{sol-exp}
\end{align}
where $\psi_S$, $\psi_{SD}$,  $\psi_D$ are the periodic Stokeslet, periodic doublet Stokeslet and periodic free dipole solutions accordingly, given by
\begin{align}
    A&=-\frac{\Omega L}{2\pi}\left(\frac{1}{2}+\ln{\frac{\pi a}{2L}}\right)\,,
 \nonumber \\
    \psi_S &= \frac{\Omega L}{4\pi} \frac{w-\bar{w}}{2i}\log\left(\sin\frac{\pi w}{2L}\right) +c.c.\,,
 \nonumber \\
    \psi_{SD} &= \frac{i \pi a^2}{L}\bar{\partial}\psi_S\,,
 \nonumber \\
    \psi_D &= \frac{i\Omega a^2}{16} \cot \frac{\pi w}{2L}+c.c.\,.
 \label{psi_solution}
\end{align}

Expanding the velocity given by (\ref{psi_solution}) in $a/L$  on the surface of the wire $w=a e^{i\alpha}$ we obtain 
\begin{align}
  &\frac{v}{\Omega L} \approx  \frac{\pi a^2 }{96 L^2}\left(4-3e^{-2i\alpha}-e^{-2i\alpha}\right) = O\left(\frac{a^2}{L^2}\right),
\end{align}
so we satisfied boundary conditions in the first two orders in $a/L$ expansion. Adding to the solution the higher-order derivatives of Stokeslet and dipole solution we can satisfy boundary conditions in any desired order. For our goals, we need only (\ref{psi_solution}).

The asymptotics of velocity at the infinity $z\to \infty$ is
\begin{align}
   v= \Omega z + \frac{\Omega L}{2\pi}\left(-\frac{1}{2}+\log{\frac{L}{\pi a}} -\frac{\pi^2 a^2}{2 L^2}+O\left(\frac{a^4}{L^4}\right)  \right).
\end{align}
Thus, for $h_{\perp}$ we obtain
\begin{align}
    h_\perp=\frac{L}{2\pi} \left(-\frac{1}{2}+\log{\frac{L }{\pi a }}-\frac{\pi^2 a^2}{2 L^2}+O\left(\frac{a^4}{L^4}\right)\right)\,.
\end{align}

\bibliographystyle{unsrt}
\bibliography{chiral-propulsion}

\begin{thebibliography}{10}

\bibitem{Zhang}
Li~Zhang, Jake~J. Abbott, Lixin Dong, Kathrin~E. Peyer, Bradley~E. Kratochvil,
  Haixin Zhang, Christos Bergeles, and Bradley~J. Nelson.
\newblock Characterizing the swimming properties of artificial bacterial
  flagella.
\newblock {\em Nano Letters}, 9(10):3663--3667, 2009.
\newblock PMID: 19824709.

\bibitem{Ghosh}
Ambarish Ghosh and Peer Fischer.
\newblock Controlled propulsion of artificial magnetic nanostructured
  propellers.
\newblock {\em Nano Letters}, 9(6):2243--2245, 2009.
\newblock PMID: 19413293.

\bibitem{Schamel}
Debora Schamel, Marcel Pfeifer, John~G. Gibbs, Björn Miksch, Andrew~G. Mark,
  and Peer Fischer.
\newblock Chiral colloidal molecules and observation of the propeller effect.
\newblock {\em Journal of the American Chemical Society}, 135(33):12353--12359,
  2013.

\bibitem{Keaveny}
Eric~E. Keaveny, Shawn~W. Walker, and Michael~J. Shelley.
\newblock Optimization of chiral structures for microscale propulsion.
\newblock {\em Nano Letters}, 13(2):531--537, 2013.
\newblock PMID: 23317170.

\bibitem{walker}
D.~Walker, M.~Kübler, K.~I. Morozov, P.~Fischer, and A.~M. Leshansky.
\newblock Optimal length of low reynolds number nanopropellers.
\newblock {\em Nano Letters}, 15(7):4412--4416, 2015.
\newblock PMID: 26030270.

\bibitem{aif2018chiral}
S~Aif, IA~Kuk, and DE~Kharzeev.
\newblock Chiral propulsion by electromagnetic fields.
\newblock {\em arXiv preprint arXiv:1804.08664}, 2018.

\bibitem{Purcell}
E.~M. Purcell.
\newblock Life at low reynolds number.
\newblock {\em American Journal of Physics}, 45(1):3--11, 1977.

\bibitem{Lighthill}
James. Lighthill.
\newblock Flagellar hydrodynamics.
\newblock {\em SIAM Review}, 18(2):161--230, 1976.

\bibitem{Liu_flag}
Bin Liu, Kenneth~S. Breuer, and Thomas~R. Powers.
\newblock Helical swimming in stokes flow using a novel boundary-element
  method.
\newblock {\em Physics of Fluids}, 25(6):061902, 2013.

\bibitem{li}
Lei Li and Saverio~E. Spagnolie.
\newblock Swimming and pumping of rigid helical bodies in viscous fluids.
\newblock {\em Physics of Fluids}, 26(4):041901, 2014.

\bibitem{Happel1983}
John Happel and Howard Brenner.
\newblock {\em Low Reynolds number hydrodynamics: with special applications to
  particulate media}.
\newblock Springer Netherlands, Dordrecht, 1983.

\bibitem{Note1}
Notice that there is no dependence of $K, C, R$ on the viscosity coefficient,
  and these tensors are purely geometric. The Stokes phenomenon can invalidate
  this statement. In the presence of the Stokes phenomenon, the inertial terms
  of the Navier-Stokes equation cannot be uniformly neglected, and the
  dissipative viscosity could appear as a cutoff.

\bibitem{Note2}
In a related approach, we could push the body through liquid with non-zero
  force and require zero-torque condition.

\bibitem{Note3}
In mechanics this symmetry results in the conservation of the combination
  $\protect \frac {d}{2\pi }P+M = const$, where $P$ is the total momentum of a
  system and $M$ is the total angular momentum \cite {LANDAU197613}.

\bibitem{shapere1987self}
Alfred Shapere and Frank Wilczek.
\newblock Self-propulsion at low reynolds number.
\newblock {\em Physical Review Letters}, 58(20):2051, 1987.

\bibitem{bazant_vinogradova_2008}
Martin~Z. Bazant and Olga~I. Vinogradova.
\newblock Tensorial hydrodynamic slip.
\newblock {\em Journal of Fluid Mechanics}, 613:125–134, 2008.

\bibitem{navier1827lois}
CLMH Navier.
\newblock Sur les lois de l’{\'e}quilibre et du mouvement des corps
  {\'e}lastiques.
\newblock {\em Mem. Acad. R. Sci. Inst. France}, 6(369):1827, 1827.

\bibitem{kamrin}
Ken Kamrin, Martin~Z. Bazant, and Howard~A. Stone.
\newblock Effective slip boundary conditions for arbitrary periodic surfaces:
  the surface mobility tensor.
\newblock {\em Journal of Fluid Mechanics}, 658:409–437, 2010.

\bibitem{Note4}
The next correction to the flow at large distances from the surface is expected
  to decay as $\protect \qopname \relax o{exp}(-kz)$, where $k=$ is a primary
  wavevector of the surface modulation.

\bibitem{LL6}
L.D. Landau and E.M. Lifshitz.
\newblock {\em Fluid Mechanics: Volume 6}.
\newblock Number v. 6. Elsevier Science, 1987.

\bibitem{davis}
A.~M.~J. Davis.
\newblock Periodic blocking in parallel shear or channel flow at low reynolds
  number.
\newblock {\em Physics of Fluids A: Fluid Dynamics}, 5(4):800--809, 1993.

\bibitem{luchini}
Paolo Luchini, Fernando Manzo, and Amilcare Pozzi.
\newblock Resistance of a grooved surface to parallel flow and cross-flow.
\newblock {\em Journal of Fluid Mechanics}, 228:87–109, 1991.

\bibitem{LANDAU197613}
L.D. Landau and E.M. Lifshitz.
\newblock Chapter ii - conservation laws.
\newblock In L.D. Landau and E.M. Lifshitz, editors, {\em Mechanics (Third
  Edition)}, pages 13 -- 24. Butterworth-Heinemann, Oxford, third edition
  edition, 1976.

\end{thebibliography}

\end{document}